\documentstyle[12pt]{article}
\begin{document}

\def\sh{\mathop{\rm sh}\nolimits}
\def\ch{\mathop{\rm ch}\nolimits}
\def\var{\mathop{\rm var}}\def\exp{\mathop{\rm exp}\nolimits}
\def\Re{\mathop{\rm Re}\nolimits}
\def\Sp{\mathop{\rm Sp}\nolimits}
\def\kp{\mathop{\text{\ae}}\nolimits}
\def\bk{{\bf {k}}}
\def\bp{{\bf {p}}}
\def\bq{{\bf {q}}}
\def\lra{\mathop{\longrightarrow}}
\def\Const{\mathop{\rm Const}\nolimits}
\def\sh{\mathop{\rm sh}\nolimits}
\def\ch{\mathop{\rm ch}\nolimits}
\def\var{\mathop{\rm var}}

\def\Re{\mbox {Re}}
\newcommand{\Z}{\mathbb{Z}}
\newcommand{\R}{\mathbb{R}}
\def\mK{\mathop{{\mathfrak {K}}}\nolimits}
\def\mR{\mathop{{\mathfrak {R}}}\nolimits}
\def\mv{\mathop{{\mathfrak {v}}}\nolimits}
\def\mV{\mathop{{\mathfrak {V}}}\nolimits}
\def\mD{\mathop{{\mathfrak {D}}}\nolimits}
\newcommand{\ccm}{{\cal M}}
\newcommand{\cE}{{\cal E}}
\newcommand{\cV}{{\cal V}}
\newcommand{\cI}{{\cal I}}
\newcommand{\cR}{{\cal R}}
\newcommand{\cK}{{\cal K}}
\newcommand{\cH}{{\cal H}}

\def\br{\mathop{{\bf {r}}}\nolimits}
\def\bS{\mathop{{\bf {S}}}\nolimits}
\def\bA{\mathop{{\bf {A}}}\nolimits}
\def\bJ{\mathop{{\bf {J}}}\nolimits}
\def\bn{\mathop{{\bf {n}}}\nolimits}
\def\bg{\mathop{{\bf {g}}}\nolimits}
\def\bv{\mathop{{\bf {v}}}\nolimits}
\def\be{\mathop{{\bf {e}}}\nolimits}
\def\bp{\mathop{{\bf {p}}}\nolimits}
\def\bz{\mathop{{\bf {z}}}\nolimits}
\def\bbf{\mathop{{\bf {f}}}\nolimits}
\def\bb{\mathop{{\bf {b}}}\nolimits}
\def\ba{\mathop{{\bf {a}}}\nolimits}
\def\bx{\mathop{{\bf {x}}}\nolimits}
\def\by{\mathop{{\bf {y}}}\nolimits}
\def\br{\mathop{{\bf {r}}}\nolimits}
\def\bs{\mathop{{\bf {s}}}\nolimits}
\def\bH{\mathop{{\bf {H}}}\nolimits}
\def\bk{\mathop{{\bf {k}}}\nolimits}
\def\be{\mathop{{\bf {e}}}\nolimits}
\def\bnul{\mathop{{\bf {0}}}\nolimits}
\def\bq{{\bf {q}}}

\newcommand{\oV}{\overline{V}}
\newcommand{\vkp}{\varkappa}
\newcommand{\os}{\overline{s}}
\newcommand{\opsi}{\overline{\psi}}
\newcommand{\ov}{\overline{v}}
\newcommand{\oW}{\overline{W}}
\newcommand{\oPhi}{\overline{\Phi}}

\def\mI{\mathop{{\mathfrak {I}}}\nolimits}
\def\mA{\mathop{{\mathfrak {A}}}\nolimits}

\def\st{\mathop{\rm st}\nolimits}
\def\tr{\mathop{\rm tr}\nolimits}
\def\sign{\mathop{\rm sign}\nolimits}
\def\d{\mathop{\rm d}\nolimits}
\def\const{\mathop{\rm const}\nolimits}
\def\O{\mathop{\rm O}\nolimits}
\def\Spin{\mathop{\rm Spin}\nolimits}
\def\exp{\mathop{\rm exp}\nolimits}

\def\mI{\mathop{{\mathfrak {I}}}\nolimits}
\def\mA{\mathop{{\mathfrak {A}}}\nolimits}

\def\st{\mathop{\rm st}\nolimits}
\def\tr{\mathop{\rm tr}\nolimits}
\def\sign{\mathop{\rm sign}\nolimits}
\def\d{\mathop{\rm d}\nolimits}
\def\const{\mathop{\rm const}\nolimits}
\def\O{\mathop{\rm O}\nolimits}
\def\Spin{\mathop{\rm Spin}\nolimits}
\def\exp{\mathop{\rm exp}\nolimits}

\centerline{\bf {THE DYNAMIC QUANTIZATION of GRAVITY}} \centerline
{\bf {and the COSMOLOGICAL CONSTANT PROBLEM}}

\centerline{ }
\centerline{\it {S.N. Vergeles \footnote{{e-mail:vergeles@itp.ac.ru}}}}
\centerline{ }
{\small {\it {
\centerline{Landau Institute for Theoretical Physics, Russian Academy of Sciences,}
\centerline{Chernogolovka, Moskow region, 142432 Russia }}}}
\centerline{ }

\vspace{3mm}
\centerline{ }

  \parbox[b]{145mm}{
 { \hspace{5mm}
{ \small { After a brief outlook of the dynamic quantization
method and application of the method to gravity the idea of
natural solution of cosmological constant problem in inflating
Universe is presented. }}}}


\centerline{ }
\centerline{ }
{\large\sl {1. Introduction }}
\centerline{ }
\centerline{ }

A new Dynamic Method for quantizing generally covariant theories
has been proposed in a series of work [1-2]. The study of
two-dimensional models, such as the two-dimensional bosonic string
[3] and two-dimensional gravity interacting with matter [4], from
the standpoint of this method has led, in the first place, to
anomaly-free quantization of these models \footnote {See [5] for a
discussion of anomaly-free quantization of two-dimensional gravity
from a standpoint close to that of the present author.} and, in
the second place, to further elaboration of the Dynamic
quantization method itself. Here we present the dynamic
quantization method and its application to four-dimensional
gravity. This is done in Sections 2 and 3. The main purpose of the
present paper is an attempt to solve the cosmological constant
problem in the framework of the theory of gravity quantized by the
dynamic method in an inflating Universe. We show in Section 4 that
 since the number of physical degrees of
freedom is finite under Dynamic quantization (in a closed model),
the quantum fluctuations in quantum Einstein equation "die out"
with time in the inflating Universe. This result is valid to all
orders in the Planck scale $l_P$.

Let's outline shortly the cosmological constant problem
\footnote{The reader can find the review of the problem in [6] }.

Consider Einstein equation with $\Lambda$-term ($\hbar=c=1$):
$$
R_{\mu\nu}-\frac12g_{\mu\nu}R=8\pi
G\,T_{\mu\nu}+\Lambda\,g_{\mu\nu}\,.  \eqno(1)
$$
Here $T_{\mu\nu}$ is energy-momentum tensor of the matter and
$\Lambda$ is some constant parameter having the dimension
[$cm^{-2}$]. In the used unit system the Newtonian gravitational
constant
$$
G\sim l_P^2\sim 2,5\cdot 10^{-66}\,{cm}^2\,,  \eqno(2)
$$
and according to experimental data the mean energy density today
is of the order
$$
T_{\mu\nu}\sim\rho_1\sim 10^8{cm}^{-4}\longrightarrow
8\pi\,G\,T_{\mu\nu}\sim 5\cdot 10^{-57}{cm}^{-2}\,,  \eqno(3)
$$
and
$$
\Lambda\sim 10^{-56}{cm}^{-2}\,.   \eqno(4)
$$
Thus, if Einstein equation (1) is used for description of the
today dynamics of Universe, the quantities in its right hand side
are of the same order indicated in (3) and (4).

Now let us estimate the possible value of the right hand side of
Eq. (1) in the framework of canonical quantum field theory. For
simplicity consider energy-momentum tensor in quantum
electrodynamics in flat spacetime:
$$
T_{\mu\nu}=-\frac{1}{4\pi}\left(F_{\mu\lambda}F_{\nu}^{\;\;\lambda}-
\frac14\eta_{\mu\nu}F^2\right)+\frac{i}{2}\left(\opsi\gamma_{(\mu}\nabla_{\nu)}\psi-
\overline{\nabla_{(\nu}\psi}\gamma_{\mu)}\psi\right)\,. \eqno(5)
$$
Casimir effect, predicted in [7] and experimentally verified in
[8], shows for reality of zero-point energies. Moreover, the
attempts to drop out zero-point energies by appropriate normal
ordering of creating and annihilating operators in energy-momentum
tensor fail for many of reasons (the discussion of this problem
see, for example, in [9]). Thus, at estimating vacuum expectation
value of energy-momentum tensor (5), it should not be performed
normal ordering of creating and annihilating operators in (5).
Thus we obtain  for vacuum expectation value of tensor (5) in free
theory:
$$
\langle T_{\mu\nu}\rangle_0=\int\frac{\d^{(3)}k}{(2\pi)^3}
\left(\frac{k_{\mu}k_{\nu}}{k^0}\bigg|_{k^0=|\bk|}-\frac{2k_{\mu}k_{\nu}}{k^0}\bigg|_{k^0=\sqrt{m^2+\bk^2}}
\right)\,. \eqno(6)
$$
Here $m$ is the electron mass. The first item in (6) gives the
positive contribution but the second item gives the negative
contribution since these items give the boson and fermion
contributions to vacuum energy, respectively. If integration in
(6) is restricted by Planck scale, $k_{max}\sim l_P^{-1}$, then
from (6) and (2) it follows:
$$
 8\pi G \langle T_{\mu\nu}\rangle_0\sim l_P^{-2}\sim
 10^{66}\,{cm}^{-2}\,.   \eqno(7)
 $$
It is clear that the interaction of fields doesn't changes
qualitatively the estimation (7). From (7) and (3) we see that the
contribution to the righthand side of Eq. (1) estimated in the
framework of canonical quantum field theory is larger about
$10^{120}$ times in comparison with the experimental estimations.

It is known that in globally supersimmetric field theories the
vacuum energy is equal to zero [10]. Indeed, in flat spacetime the
anticommutation relations
 $$
 \{Q_{\alpha},\,Q^{\dag}_{\beta}\}=(\sigma_{\mu})_{\alpha\beta}{\cal
 P}^{\mu}\,.   \eqno(8)
 $$
 take place. Here $Q_{\alpha}$ are supersimmetry generators,
 $\alpha$ è $\beta$ are spinor indexes,
 $\sigma_1,\,\sigma_2$ and $\sigma_3$ are the Pauli matrices, $\sigma_0=1$, and ${\cal
 P}^{\mu}$ is the energy-momentum 4-vector operator.
 If supersimmetry is unbroken, then the vacuum state
 $|0\rangle$ satisfies
  $$
Q_{\alpha}|0\rangle=Q_{\alpha}^{\dag}|0\rangle=0\,,   \eqno(9)
$$
and (8) and (9) imply
$$
\langle {\cal P}^{\mu}\rangle_0=0\,.   \eqno(10)
$$
The equality (10) means that the total sum of zero-point energies
in unbroken globally supersimmetric field theories is rigorously
equal to zero.

However, even if supersimmetry takes place on fundamental level,
it is broken on experimentally tested scales. If one assumes that
supersimmetry is unbroken on the scales greater than $k_{SS}\sim
10^{17}\,cm^{-1}\,(\sim 10^3\,GeV)$, then even in this case the
contribution to the right hand side of Eq. (1) from zero-point
energies of all normal modes with energies less then $k_{SS}$ will
exceed experimentally known value about $10^{58}$ times.

It follows from the said above that any calculation in the
framework of canonical quantum field theory leads to unacceptable
large vacuum expectation value of energy momentum tensor. The
considered catastrophe isn't solved at present in superstring
theory.

It should be noted here that the problem of cosmological constant
is solved in original theory of G. Volovik [11]. In this theory
the gravitons and other excitations are the quasiparticles in a
more fundamental quantum system --- quantum fluid of the type
${}^3He$ in superfluid phase. Another approach to the problem of
cosmological constant in the frame of M-theory is developed in
works [14]. Interesting idea about solution of the problem is
presented in [15].

Specific results of the application of the Dynamic quantization
method to the above mentioned two-dimensional theories [3, 4],
obtained by explicit constructions and direct calculations,
justify the abstract assumptions and axioms on which this method
is based.

We shall explain the ideology and logical scheme of the Dynamic
method taking account of the experience in quantizing
two-dimensional gravity.

The key point in the quantization of two-dimensional gravity was
the construction of a complete set of such operators $ \{A_n,
B_n,\ldots \} $, designated below as $\{A_N, \,A_N^{\dag}\}$, which
possess the following properties:

\centerline{}

1) The operators $A_N$ and $A_N^{\dag}$ are Hermithian conjugates
of one another and
$$
[\,A_N,\,A_M\,]=0, \qquad [A_N,\,A^{\dag}_M\,]=\delta_{NM}\,.
\eqno(11)
$$

2) The set of operators $\{A_N,\,A_N^{\dag}\}$ describes all
physical dynamical degrees of freedom of a system.

3) Each operator from the set $\{A_N,\,A_N^{\dag}\}$ commutes
weakly with all first class constraints or with the complete
Hamiltonian of the theory.

\centerline{}

Quantization is performed directly using the operators
$\{A_N,\,A_N^{\dag}\}$. It means that the space of physical states
is created using the operators $\{A_N^{\dag}\}$ from the ground
state and all operators are expressed in terms of the operators
$\{A_N,\, A_N^{\dag}\}$, as well as in terms of the operators
describing the gauge degrees of freedom. However, in the theory of
two-dimensional gravity the operators
 $\{A_N,\,A_N^{\dag}\}$ were constructed explicitly (i.e., they were
 expressed explicitly in terms of the fundamental dynamical variables),
in more realistic theories this problem is hardly solvable.
Therefore, the set of operators $\{A_N,\,A_N^{\dag}\}$ with
properties 1)--3) must be introduced axiomatically. Conversely, the
properties 1)--3) make it possible, in principle, to express the
initial variables in terms of the convenient operators
$\{A_N,\,A_N^{\dag}\}$.

However, in contrast to the two-dimensional theory of gravity,
regularization is necessary in real models of gravity. In the
Dynamic quantization method, regularization is carried out
precisely in terms of the operators $\{A_N,\,A_N^{\dag}\}$. As will
be shown below, such regularization is natural in generally
covariant theories, since it preserves the form of the Heisenberg
equations and thereby also the general covariance of the theory.

As a result we have the regularized general covariant theory
describing quantum gravity, the main property of which is the
finiteness of physical degrees of freedom contained in each finite
volume. Evidently, the theory of discrete quantum gravity defined
on the lattice (simplicial complexes) possess the same properties
(see, for example, [12] and references here). Therefore, one can
think that the theory of gravity quantized by dynamic quantization
method is the continuous limit of discrete quantum gravity
discussed in [12].

\centerline{ }
\centerline{ }
{\large\sl {2. Method of dynamic quantization }}
\centerline{ }
\centerline{ }

Let's consider a generally covariant field theory. Let us assume
that in this theory the Hamiltonian in the classical limit is an
arbitrary linear combination of the first class constraints and
there are no the second class constraints.

Let $\{\Phi^{(i)}(x),\,P^{(i)}(x)\}$ be a complete set of
fundamental fields of the theory and their canonically conjugate
momenta, in terms of which all other physical quantities and
fields of the theory are expressed. Here the index $(i)$
enumerates the tipe of fields. For example, for some $(i)$ these
can be either 6 spatial components of the metric tensor
$g_{ij}(x)$ or the scalar field $\phi(x)$ or the Dirac field
$\psi(x)$ etc. The set of fields $\{\Phi^{(i)}(x)\}$ is a complete
set of the mutually commuting (at least in formal unregularized
theory) fundamental fields of the theory.

Next, to simplify the notation the index $i$ will be omitted. It
can be assumed that the variable $x$ includes, besides the spatial
coordinates, the index $i$ also.

The construction of a quantum theory by the Dynamic method is
based on the following natural assumptions relative to the
structure of the space $F$ of the physical states of
the theory. \\

{\bf {1.}} {\it {All states of the theory having physical sense
are obtained from the ground state}} $\vert \,0\,\rangle$ {\it
{using the creation operators}} $A^{\dag}_N$:
 $$
 \vert \,n_1,\,N_1;\ldots
 ;\,n_s,\,N_s\,\rangle=(n_1!\cdot\ldots\cdot n_s!\,)^{-\frac{1}{2}} \cdot
 (A^{\dag}_{N_1})^{n_1}\cdot\ldots\cdot
 (A^{\dag}_{N_s})^{n_s}\,\vert \,0\,\rangle \ ,
 $$
 $$
   A_N\,\vert\,0\,\rangle=0 \,.                              \eqno(12)
 $$
{\it {States (12) form an orthonormal basis of the space $F$ of
physical states of the theory.}} \\

The numbers $n_1,\ldots,n_s$ assume integer values and are called
occupation numbers. \\

{\bf {2.}} {\it {The set of states}}
  $\Phi(x)\,\vert\,n_1,N_1;\ldots;n_s,N_s\,\rangle$,
  {\it {, where the set of numbers $(n_1, N_1; \ldots; n_s, N_s)$ is
fixed, contains a superposition of {\bf {all}} states of the
theory, for which one of the occupation number differs in modulus
by one and all other occupation numbers equal to the occupation
numbers of state (12).}} \\

Here the operators $A^{\dag}_N$ and their conjugates $A_N$ possess
the standard commutation properties (11). The operators
$\{A_N,\,A_N^{\dag}\}$, generally speaking, can be bosonic or
fermionic. If the creation and annihilation operators follow the
Fermi statistics, then the anticommutator is taken in (11). For
the case of compact spaces which is interesting for us, we can
assume without loss of generality that the index $N$, enumerating
the creation and annihilation operators, belongs to a discrete
finite-dimensional lattice. A norm can be easily introduced in the
space of indexes $N$.

Since states (12) are physical, they satisfy the relations
$$
{\cal H}_T \, \vert \,n_1,\,N_1;\ldots;\,n_s,\,N_s\,\rangle=0\,\,,
\eqno(13)
$$
where ${\cal H}_T$ is the complete Hamiltonian of the theory. We
assume that ${\cal H}_T=\sum_{\Xi}v_{\Xi}\chi_{\Xi}$, where
$\{\chi_{\Xi}\}$ is the complete set of the first class
constraints and $\{v_{\Xi}\}$ is arbitrary set of Lagrange
multipliers.

Equations (12) and (13) are compatible if and only if the
following relations are valid:
$$
[A_N,\,\chi_{\Xi}]=\sum_{\Pi}c_{N\Xi\Pi}\,\chi_{\Pi}\,,
$$
$$
[A_N^{\dag},\,\chi_{\Xi}]=-\sum_{\Pi}\chi_{\Pi}\,c_{N\Xi\Pi}^{\dag}=
\sum_{\Pi}\tilde{c}_{N\Xi\Pi}\,\chi_{\Pi}\,,\eqno(14)
$$
Since the coefficients $c_{N\Xi\Pi}, \; \tilde{c}_{N\Xi\Pi}$ in
Eq. (14) generally are operators, the arrangement of the
multiplies in the right hand sides of Eqs. (14) is important.

Let $(A^{\dag}_N,\,A_N)$ be a pair of bose or fermi creation and
annihilation operators creating or annihilating the state with the
wave function $\psi_N(x)$. According to (14) we have:
$$
[A_N,\,{\cal
H}_T]=\sum_{\Xi,\,\Pi}r_{N\,\Xi\,\Pi}v_{\Xi}\,\chi_{\Pi}
\longleftrightarrow [A_N^{\dag},\,{\cal
H}_T]=-\sum_{\Xi,\,\Pi}\chi_{\Pi}\,v_{\Xi}^*r_{N\,\Xi\,\Pi}^{\dag}\,.
\eqno(15)
$$
Let an arbitrary operator $\Phi$ be represented as a normal
ordered power series in operators $(A^{\dag}_N,\,A_N)$:
$$
\Phi=\Phi'+\phi_NA_N+A^{\dag}_N\tilde{\phi}_N\,,  \eqno(16)
$$
By definition here the operator $\Phi'$ does not depend on the
operators $(A^{\dag}_N,\,A_N)$:
$$
[\Phi',\,A^{\dag}_N]=[\Phi',\,A_N]=0\,.  \eqno(17)
$$
It follows from Eqs. (15)-(17) that
$$
[\Phi,\,{\cal H}_T]=[\Phi',\,{\cal
H}'_T]+\sum_{\Xi}(q_{\Xi}\chi_{\Xi}+\chi_{\Xi}\tilde{q}_{\Xi})+(p_NA_N+A_N^{\dag}\tilde{p}_N)\,.
\eqno(18)
$$
Here the total Hamiltonian ${\cal H}_T$ is represented according
to (16), so that ${\cal H}'_T$ does not depend on the operators
$(A^{\dag}_N,\,A_N)$. To verify Eq. (18) let's write out the
following chain of equalities:
$$
[\Phi,\,{\cal H}_T]=[\Phi',\,{\cal H}_T]+\left(\phi_N[A_N,\,{\cal
H}_T]+[A_N^{\dag},\,{\cal H}_T]\,\tilde{\phi}_N\right)+
\left([\phi_N,\,{\cal H}_T]\,A_N+A_N^{\dag}[\tilde{\phi}_N,\,{\cal
H}_T]\right)\,.    \eqno(19)
$$
As a consequence of Eqs. (15) the second item in the right hand
side of Eq. (19) has the same structure as the second item in the
right hand side of Eq. (18). Evidently, the last items in the
right hand side of Eq. (19) has the same structure as the last
items in the right hand side of Eq. (18). Now let's write out the
following identity:
$$
[\Phi',\,{\cal H}_T]\equiv[\Phi',\,{\cal H}'_T]+[\Phi',\,{\cal
H}_T-{\cal H}'_T]\,.   \eqno(20)
$$
By definition
$$
{\cal H}_T-{\cal H}'_T=h_N\,A_N+A^{\dag}_N\,\tilde{h}_N\,.
\eqno(21)
$$
It follows from (17) and (21) that
$$
[\Phi',\,{\cal H}_T-{\cal
H}'_T]=[\Phi',\,h_N]\,A_N+A^{\dag}_N\,[\Phi',\,\tilde{h}_N]\,.
\eqno(22)
$$
Combining Eqs. (19), (20) and (22) we come to the Eq. (18).

Now let's impose an additional pair of second class constraints
$$
A_N=0\,, \qquad A^{\dag}_N=0\,.   \eqno(23)
$$
By definition under the constraints (23) any operator $\Phi$ is
reduced to the operator $\Phi'$ in (16). The Dirac bracket arising
under the constraints (23) is defined according to the following
equality:
$$
[\Phi,\,{\cal F}]^*\equiv[\Phi',\,{\cal F}']\,.  \eqno(24)
$$
The remarkable property of the considered theory is the fact that
$$
[\Phi,\,{\cal H}_T]^*\approx [\Phi,\,{\cal H}_T]\,.  \eqno(25)
$$
Here the approximate equality means that after the imposition of
all first and second class constraints the operators in the both
sides of Eq. (25) coincide, that is the weak equality (25) reduces
to the strong one. Relation (25) follows immediately from Eqs.
(15) and (24). Eq. (25) means that the Heisenberg equation
$$
i\dot{\Phi}=[\Phi,\,{\cal H}_T]^*
$$
for any field in reduced theory coincides weakly with
corresponding Heisenberg equation in nonreduced theory. Evidently,
this remarkable conclusion retains true under imposition of any
number of pairs of the second class constraints of type
(23)\footnote{In Appendix we give the simple example in which the
imposition of second class constraints of type (23) does not
change equations of motion.}.

The above-stated bring to the following idea of ultraviolet
regularization of quantum theory of gravity. Let a local field
$\Phi(x)$ create and annihilate particles in the states with wave
functions $\{\phi_N(x)\}$ by creation and annihilation operators
$\{A^{\dag}_N,\,A_N\}$ (for simplicity the field $\Phi$ is assumed
to be real). The physical space of states is invariant relative to
the action of creation and annihilation operators. Therefore there
is the possibility of imposing the second class constraints of the
type (23) for any number of pairs of these operators without
changing Heisenberg equations of motion. Let the high-frequency
(in some sense) wave functions $\{\phi_N(x)\}_{|N|>N_0}$ have the
value of index $|N|>N_0$. The ultraviolet regularization of the
theory is performed by imposing the constraints of the type (23)
for all $|N|>N_0$. It is very important that under the constraints
the regularized equations of motion and first class constraints
preserve their canonical form. Hence the equations of regularized
theory are general covariant, i.e. they conserve their form under
the general coordinate transformations and local frame
transformations.

Since unregularized theory of quantum gravity is mathematically
meaningless, so it seems correct the direct definition of
regularized theory by means of introduction of natural axioms. \\

{\bf {Axiom 1.}} {\it {All states of the theory which are
physically meaningful are obtained from the ground state}} \
$\vert \, 0 \,\rangle $ \ {\it {using the creation operators}} \
$A^{\dag}_N$ \ with \ $\vert \, N \,\vert <N_0$ \ :
 $$
 \vert \,n_1,\,N_1;\ldots ;\,n_s,\,N_s\,\rangle=(n_1!\cdot\ldots\cdot
 n_s!\,)^{-\frac{1}{2}} \cdot
 (A^{\dag}_{N_1})^{n_1}\cdot\ldots\cdot
 (A^{\dag}_{N_s})^{n_s}\,\vert \,0\,\rangle \ ,
 $$
 $$
   A_N\,\vert\,0\,\rangle=0\,.
  \eqno(26)
 $$
{\it{States (26) form an orthonormal basis of the space \
$F$ \ of physical states of the theory.}} \\

{\bf{Axiom 2.}} {\it{The dynamical variables}} \ ${\Phi}(x)$ \
{\it {transfer state (26) with fixed values of numbers $(n_1, \,
N_1; \ldots; \, n_s, \, N_s)$ into a superposition of the states
of the theory of form (26), containing all states in which one of
the occupation numbers is different in modulus by one and all
other
occupation numbers are identical to those of state (26).}} \\

{\bf{Axiom 3.}} {\it{The equations of motion and constraints for
the physical fields \ $\{\Phi (x), \, {\cal P} (x) \}$ have the
same form, to within the arrangement of the operators, as the
corresponding classical equations and constraints.}} \\

Further we suppose that the momentum variables ${\cal P} (x)$ are
expressed through the fundamental field variables $\Phi (x)$ and
their time derivatives $\dot{\Phi}(x)$, so that the Lagrange
equations instead of Hamilton equations are used.

Let's assume, further, that the ground state $\vert \, 0 \,\rangle$
is a coherent state with respect to the gauge degrees of freedom.
It means that the quantum fluctuations of the gauge degrees of
freedom are not significant and their dynamics in fact is
classical.

Let's emphasize that this assumption is related with the fact of
noncompactness of the gauge group. (Since the group of general
linear transformations is noncompact, so the gauge group in the
theory of gravity is noncompact.) In this relation it is
interesting to demonstrate the transformation of quantum particle
to classical one in the course of time in the case of noncompact
dynamics. Let $x$ and $p$ be Heisenberg coordinate and momentum
operators of a free nonrelativistic particle with mass \ $m$
moving in noncompact flat space. Then
$$
p=p_0\,, \qquad x=x_0+\frac{p_0}{m}\,t\,,
$$
where \ $t$ \ is the time, and $x_0$ and $p_0$ are constant
operators, satisfying the commutation relation $[x_0,
p_0\,]=i\,\hbar$. It is obvious that if $\langle\,p_0\,\rangle\neq
0$, then as $t\rightarrow\infty$
$$
\frac{\langle\,x\,\rangle\,\langle\,p\,\rangle}
{\mid\langle\,[\,x\,,\,p\,]\,\rangle\mid } \rightarrow\infty
\,\,,
$$
which means that the dynamics of the free particle in the course of
time becomes quasiclassical.

The quasiclassical charecter of dynamics of gauge degrees of
freedom seems true only for noncompact gauge groups. On the
contrary, the motion in compact gauge group (such as in Yang-Mills
theory) can not be regarded as classical.

Let's consider, for example, the quantized electrodynamic field in
noncovariant Coulomb gauge. In this gauge only the degrees of
freedom describing photons fluctuate, but the gauge (longitudinal)
degrees of freedom are defined unambiguously through the electric
current. Thus, the gauge degrees of freedom in QED does not
fluctuate, effectively they are classical. On the other hand, in
high-temperature confinement phase in QED on a lattice the
high-temperature expansion is valid. In this case the gauge
degrees of freedom can not be regarded as classical. So our
assumption about classical behavior of gauge degrees of freedom in
quantum gravity is equivalent to the assumption that quantum
gravity is in noncompact phase.

Consider any fundamental field:
$$
\Phi (x)=\Phi_{(cl)} (x)+ \sum_{|N|<N_0} \,[\,\phi_N
(x)\,A_N+\phi^*_N (x)\,A^{\dag}_N\,]+\ldots \,. \eqno(27)
$$
{\it {On the right-hand side of Eq.}} (27) {\it {all functions \
$\Phi_{(cl)} (x),\,\, \phi_N (x)$, \ and so on are $c$-number
functions}}.

This follows from the assumption about the quasiclassical
character of the dynamics of gauge degrees of freedom.

Now we can supplement our system of axioms by the following
supposition: field (27) is used in axioms 1-3.
 The fields $\Phi_{(cl)} (x),\,\phi_N (x),\,\psi_N (x)$,
and so on satisfy certain equations which can be obtained uniquely
from the Lagrange equations of motion, if the expansion of the
field $\Phi (x)$ in form (27) is substituted into them and then,
after normal ordering of the operators $\{\,A_N,\,A^{\dag}_N\,\}$,
the coefficients of the various powers of the generators of the
Heisenberg algebra $\{\,A_N,\,A^{\dag}_N\,\}$ are equated to zero.
As a result of the indicated normal ordering, the relations arise
between the higher order coefficient functions and the lower order
coefficient functions in expansion (27). We obtain an infinite
chain of equations for the coefficient functions $\{\,\Phi^{(cl)}
(x),\,\phi_N (x), \,\psi_N (x),\,\ldots\}$. The latter conjecture
can be
introduced with the aid of the following axiom, replacing axiom 3. \\

{\bf {Axiom $3^{\prime}$.}} {\it {The equations of motion for the
quantized fields}} (26), {\it {up to the ordering of the quantized
fields, have the same form as the corresponding classical
equations
of motion.}} \\

\centerline{ }
\centerline{ }
{\large\sl {3. Dynamic quantization
of gravity }}
\centerline{ }
\centerline{ }

We shall now apply the quantization scheme developed above to the
theory of gravity.

Let's consider the theory of gravity with a $ \Lambda $ term which
is coupled minimally with the Dirac field. The action of such a
theory has the form
$$
S=-\frac{1}{l^2_P}\int d^4\,x\,\sqrt{-g}\,\,(R+2\Lambda)+
$$
$$
+\int d^4\,x\,\sqrt{-g}\,\,\biggl\{\frac{i}{2}\,e^{\mu}_{a} \Bigl(
\overline{\psi}\gamma^a\,D_{\mu}\psi-\overline{D_{\mu}\psi}\,
\gamma^a\,\psi\Bigr)-m\overline{\psi}\psi\biggr\} \eqno(28)
$$
Here $\{e^{\mu}_a\}$ is an orthonormalized basis, \ $g_{\mu\nu}$ \
is the metric tensor, and \ $\eta_{ab}=diag(1,\,-1,\,-1,\,-1)$ \,
so that
$$
g_{\mu\nu}\,e^{\mu}_a\,e^{\nu}_b=\eta_{ab}, \ \ \ \
R=e^{\mu}_a\,e^{\nu}_b\,R^{ab}_{\mu\nu}\,,
$$
the 2-form of the curvature is given by
$$
d\omega^{ab}+\omega^a_c\wedge\omega^{cb}=
 \frac{1}{2}\,R^{ab}_{\mu\nu}\,dx^{\mu}\wedge dx^{\nu}\,,
$$
where the 1-form \ $\omega^a_b =\omega^a_{b\mu}\,dx^{\mu}$ \ is the
connection in the orthonormal basis \ $\{e^{\mu}_a\}$. The spinor
covariant derivative is given by the formula
$$
D_{\mu}\psi=\left(\frac{\partial}{\partial x^{\mu}}+
\frac{1}{2}\,\omega_{ab\mu}\,\sigma^{ab}\right)\,\psi\,,  \ \
\sigma^{ab}=\frac{1}{4}\,[\gamma^a,\,\gamma^b]\,,
$$
$\gamma^a$  are the Dirac matrices:
$$
\gamma^a\,\gamma^b+\gamma^b\,\gamma^a=2\,\eta^{ab}\,.
$$

Let's write out the equations of motion for system (28).

The variation of action (28) relative to the connection gives the equation
$$
\nabla_{\mu}\,e^a_{\nu}-\nabla_{\nu}\,e^a_{\mu}=-\frac{1}{4}l^2_P\,
\varepsilon^a_{\,\,bcd}\,e^b_{\mu}e^c_{\nu}\,\overline{\psi}\gamma^5
\gamma^d\,\psi\equiv T^a_{\mu\nu}\,.                         \eqno(29)
$$
In deriving the last equation, we employed the equality
$$
\gamma^a\,\sigma^{bc}+\sigma^{bc}\,\gamma^a=
-i\varepsilon^{abcd}\,\gamma^5\,\gamma_d        \eqno(30)
$$
Here \ $\varepsilon_{abcd}$ \ is the absolutely antisymmetric
tensor, and \ $ \varepsilon_{0123}=1$. The right-hand side of Eq.
(29) is the torsion tensor.

We note that torsion (29) possesses the property
$$
T^{\nu}_{\mu\nu}\equiv e^{\nu}_a\,T^a_{\mu\nu}\equiv 0
 \eqno(31)
$$
Consequently, even though torsion exists in the considered theory,
the torsion tensor is not present in the Dirac equation:
$$
\bigl(i\,e^{\mu}_a\,\gamma^a\,D_{\mu}-m\bigr)\,\psi=0
    \eqno(32)
$$

The variation of action (28) relative to the orthonormal basis
gives the Einstein equation, which we write in the form
$$
R_{\mu\nu}+\Lambda\,g_{\mu\nu}=\frac{1}{2}l^2_P\,\biggl\{
\frac{i}{2}\Bigl(\overline{\psi}\,\gamma^c\,e_{c(\mu}
D_{\nu)}\psi- e_{c(\mu}\overline{D_{\nu
)}\psi}\,\gamma^c\,\psi\Bigr)-
\frac{1}{2}m\,\overline{\psi}\psi\,g_{\mu\nu}\biggr\} \eqno(33)
$$
Here the expression in braces is $(T_{\mu\nu}-1/2\,g_{\mu\nu}\,T)$,
where $T_{\mu\nu}$ is the energy-momentum tensor on the mass shell
(i.e., taking account of the equations of motion of matter --- in
our case, the Dirac equation (31)).

Equations (29), (30-33), together with the relations
$$
g_{\mu\nu}=\eta_{ab}e^a_{\mu}e^b_{\nu}\,,
\ \ \ e^{\mu}_a\,e^b_{\mu}=\delta^b_a
$$
form a complete system of classical equations of motion and
constraints for system (28).

We now represent the fields as the sum of classical and quantum
components:
$$
g_{\mu\nu}=g_{(cl)\mu\nu}+h_{\mu\nu}\,\,\,, \,\,\, e^a_{\mu}=
e^a_{(cl)\mu}+f^a_{\mu}                      \eqno(34)
$$
Assume that the fermion field has no classical component, so that
$$
\psi (x)=\sum_{|N|<N_F} \Bigl(a_N\,\psi_N^{(+)}(x)+b_N^{\dag}
\,\psi_N^{(-)}(x)\Bigr)+\ldots \,\,\,,          \eqno(35)
$$
where the Fermi creation and annihilation operators satisfy the
following anticommutation relations (as usual, only the nonzero
relations are written out):
$$
\{\,a_M\,,\,a_N^{\dag}\,\}=\{\,b_M\,,\,b_N^{\dag}\,\}=
\delta_{M,N}\,, \qquad a_N|0\rangle=b_N|0\rangle=0\,. \eqno(36)
$$
The complete orthonormal set of fermion modes
$\Bigl\{\psi_N^{(\pm)}(x)\Bigr\}$ can be naturally determined as
follows. Denote by $\Sigma^{(3)}$ the spacelike hypersurface,
defined by the equation $t=\Const$, and by $\Sigma^{(3)}_0$ the
hypersurface at $t=t_0$. Let the metric in space-time be given by
means of the tensor $g_{\mu\nu}$. This metric induces a metric on
$\Sigma^{(3)}_0$, which in the local coordinates $x^i\,,\,i=1,2,3,$
is represented by the metric tensor ${}^3g _{ij}$. Using the
equations
$$
{ }^3g_{ij,k}=\gamma^l_{ik}\,{ }^3g_{lj}+\gamma^l_{jk}\,{ }^3g_{il}\,,
 \ \ \ \gamma^k_{ij}=\gamma^k_{ji}\,,
$$
$$
{ }^3g_{ij}=-\sum^3_{\alpha=1}\,{ }^3e^{\alpha}_i\,{ }^3e^{\alpha}_j\,,
\ \ \ { }^3e^{\alpha}_i\,{ }^3e^i_{\beta}=\delta_{\alpha\beta}\,,
$$
$$
\partial_i{ }^3e^i_{\alpha}+\gamma^j_{ki}\,{ }^3e^k_{\alpha}+
{ }^3\omega_{\alpha\beta i}\,{ }^3e^j_{\beta}=0\,, \ \ \
{ }^3\omega_{\alpha\beta i}=-{ }^3\omega_{\beta\alpha i}
$$
the connection (without torsion) \ $\gamma^i_{jk}$ \ in local
coordinates and a spin connection \ ${ }^3\omega_{\alpha\beta i}$
are determined on \ $\Sigma^{(3)}_0$. For a Dirac single-particle
Hamiltonian we have:
$$
{\cal H}_D=-i\,{ }^3e_{\alpha}^i\,\alpha^{\alpha}\,
(\partial_i+\frac{1}{2}\,{ }^3\omega_{\beta\gamma
i}\,\frac{1}{4}\,
[\alpha^{\beta},\,\alpha^{\gamma}\,]\,)+m\,\gamma^0\,,
$$
$$
\alpha^{\beta}=\gamma^0\,\gamma^{\beta}
$$
It is easy to check that in the metric
$$
\langle\,\psi_M,\,\psi_N\,\rangle=\int_{\Sigma^{(3)}_0}\,d^3x\,
\sqrt{-{ }^3g}\,\psi^{\dag}_M\,\psi_N   \eqno(37)
$$
the operator \ ${\cal H}_D$ \ is self-conjugated. Consequently,
the solution of the problem for the eigenvalues on \
$\Sigma^{(3)}_0$
$$
{\cal
H}^{(0)}_D\,\psi^{(\pm)}_N(x)=\pm\varepsilon_N\,\psi_N^{(\pm)}(x)\,,
\ \ \ \
 \varepsilon_N > 0   \eqno(38)
$$
has a complete set of orthonormal modes in metric (37). The index
(0) everywhere means that in the corresponding quantity the fields
are taken in the zero approximation with respect to quantum
fluctuations.

Note that a one-to-one relation can be established between the
positive- and negative-frequency modes by means of the equation
$$
\gamma^0\gamma^5\,\psi^{(+)}_M=\psi_M^{(-)}
$$

We call the attention to the fact that the scalar product
$$
(\psi_M,\,\psi_N)=\int_{\Sigma^{(3)}}\,d^3x\,\sqrt{-g^{(0)}}\,
\psi^{\dag}_M\,\psi_N    \eqno(39)
$$
is not always the same as the scalar product (37). These scalar
products coincide, if the path function \ $N=1$, which happens, for
example, for the metric
$$
g^{(0)}_{0i}=0\,, \ \ \ g^{(0)}_{00}=1\,.
$$
The scalar product (39) has the advantage over the scalar product
(37) that if the modes \ $\{\psi^{(\pm)}_N(x)\,\}$ \ satisfy the
Dirac equation in the zero approximation with respect to quantum
fluctuations (which, according to the exposition below, does indeed
happen), then the scalar product (39) is conserved in time.

The field $h_{\mu\nu}$ in Eq.(34) can be expanded as follows:
$$
h_{\mu\nu}= l_P \sum_{|N|<N_0}(h_{N\,\,\mu \nu} c_N+ h^*_{N\,\,\mu
\nu} c^{\dag}_N)+
$$
$$
+l^2_P \Bigl\{\sum_{|N_1|,|N_2|<N_0} (h_{N_1 N_2\,\,\mu
\nu}c_{N_1} c_{N_2}+ h^*_{N_1 N_2\,\,\mu \nu}c^{\dag}_{N_1}
c^{\dag}_{N_2}+ h_{N_1 \mid N_2\,\,\mu \nu}c^{\dag}_{N_1}
c_{N_2})+
$$
$$
+\sum_{|N_1|,\,|N_2|<N_F}\,(\,h^{F(++)}_{N_1\,N_2\,\mu\nu}\,
a^{\dag}_{N_1}\,a_{N_2}+
h^{F(--)}_{N_2\,N_1\,\mu\nu}\,b^{\dag}_{N_1}\,b_{N_2}+
$$
$$
+h^{F(+-)}_{N_1\,N_2\,\mu\nu}\,a^{\dag}_{N_1}\,b^{\dag}_{N_2}+
h^{F(+-)*}_{N_1\,N_2\,\mu\nu}\,b_{N_2}\,a_{N_1})\Bigr\}+\ldots
\eqno(40)
$$
In Eqs.(34), (35), and (40) the $c$-number coefficient fields \
$\psi^{(\pm)}_N\,,\,g_{(cl)\,\mu\nu}$ , \ $h_{N\,\mu\nu}$ \ and so
on can be expanded in powers of the Planck scale, for example
$$
g_{(cl)\,\mu\nu}=g^{(0)}_{\mu\nu}+l^2_p\,g^{(2)}_{(cl)\,\mu\nu}+\ldots
$$
Since fields (40) are real, we have
$$
h_{N_1 N_2\,\,\mu \nu}=h_{N_2 N_1\,\,\mu \nu}\,,\,\,
h^*_{N_1 \mid N_2\,\,\mu \nu}=h_{N_2 \mid N_1\,\,\mu \nu}\,,
$$
$$
h^{F(++)*}_{N_2\,N_1\,\mu\nu}=h^{F(++)}_{N_1\,N_2\,\mu\nu}\,,\,\,
h^{F(--)*}_{N_2\,N_1\,\mu\nu}=h^{F(--)}_{N_1\,N_2\,\mu\nu}
  \eqno(41)
$$
The operators \ $\{c_N,\,c_N^+\,\}$ \ satisfy the Bose commutation
relations (31). A method for choosing the set of functions \
$\{h_{N\,\mu\nu}\,\}$ \ will be discussed below.

According to the dynamic quantization scheme, we must substitute
fields (34-35) and (40) into Eqs. (29) and (32-33), after which the
operators $\{\,A_N,\,A^{\dag}_N\,\}$ must be normal-ordered and all
coefficients of the various powers of these operators and the
Planck scale must be set equal to zero.

Thus, we obtain the first of these equations:
$$
\nabla^{(0)}_{\mu}\,e^{(0)\,a}_{\nu}-\nabla^{(0)}_{\nu}\,e^{(0)\,a}_{\mu}=0\,,
\,\,\,
R^{(0)}_{\mu \nu}+\Lambda\,g^{(0)}_{\mu \nu}=0          \eqno(42)
$$
Here and below all raising and lowering of indices are done with
the tensors $g^{(0)}_{\mu \nu}$ and $g^{(0)\mu\nu}$. Thus, in the
lowest approximation the fields satisfy the classical equations of
motion. In the zeroth approximation we also have a series of
equations for the fermion modes:
$$
\bigl(i\,e^{(0)\,\mu}_a\,\gamma^a\,D^{(0)}_{\mu}-m\,\bigr)\,
\psi^{(0)(\pm)}_N=0    \eqno(43)
$$

We now introduce the notation
$$
K^{(0) \lambda \rho}_{\mu \nu}=\left [-\frac{1}{2}\nabla^{(0)}_{\sigma}
\nabla^{(0)\sigma}\,\delta^{\lambda}_{\mu}\,\delta^{\rho}_{\nu}-
R^{(0)\lambda\,\,\rho}_{\,\,\,\mu\,\,\,\nu}+
R^{(0)\rho}_{\nu}\,\delta^{\lambda}_{\mu}+\right.
$$
$$\left.+\nabla^{(0)}_{\mu}\,\left (\,\nabla^{(0)\lambda}\,
\delta^{\rho}_{\nu}-\frac{1}{2}\,\nabla^{(0)}_{\nu}\,g^{(0)\lambda \rho}
\right )\,\right ]+[\,\mu \longleftrightarrow \nu ]+
2\,\Lambda\,\delta^{\lambda}_{(\mu}\delta^{\rho}_{\nu)}\,,
                                                 \eqno(44)
$$
$$
R^{(0)(2)}_{\mu \nu}(h,\,h^{\prime})=\frac{1}{2}\,\left [\,R^{(0)(2)}_{\mu \nu}
(h+h^{\prime},\,h+h^{\prime})-R^{(0)(2)}_{\mu \nu}(h,\,h)-
R^{(0)(2)}_{\mu \nu}(h^{\prime},\,h^{\prime}) \right ]       \eqno(45)
$$
It is easily checked that
$$
\frac{1}{2}\,K^{(0)\,\lambda\rho}_{\mu\nu}=\frac{\delta\,
(R_{\mu\nu}+\Lambda\,g_{\mu\nu})}{\delta\,g_{\lambda\rho}}\,\bigg|_{g_{\mu\nu}=
g^{(0)}_{\mu\nu}}\,,
$$
where $R^{(0)(2)}_{\mu\nu}(h,\,h)$ is a quadratic form of the
tensor field $h_{\lambda\rho}$ which can be constructed in terms of
the second variation of \ $R_{\mu\nu}$ relative to the metric
tensor at the point \ $g^{(0)}_{\mu\nu}$. Let's write out the
complete form:
$$
R^{(0)(2)}_{\mu\nu}(h,h)=
\frac{1}{2}\,(h^{\rho}_{\lambda}\,
h^{\lambda}_{\rho ;\mu}\,)_{;\nu}-
\frac{1}{2}\,[\,h^{\lambda}_{\sigma}\,(
h^{\sigma}_{\mu;\nu}+h^{\sigma}_{\nu;\mu}-
h^{;\sigma}_{\mu\nu}\,)\,]_{;\lambda}+
$$
$$
+\frac{1}{4}\,h^{\lambda}_{\lambda;\rho}\,
(h^{\rho}_{\mu;\nu}+h^{\rho}_{\nu;\mu}-
h^{;\rho}_{\mu\nu}\,)-
\frac{1}{4}\,(h^{\lambda}_{\rho;\nu}+
h^{\lambda}_{\nu;\rho}-h^{;\lambda}_{\nu\rho}\,)
\,(h^{\rho}_{\mu;\lambda}+
h^{\rho}_{\lambda;\mu}-h^{;\rho}_{\mu\lambda}\,)
$$
Thus, $R^{(0)(2)}_{\mu\nu}(h,\,h^{\prime})$ \ is a symmetric
bilinear form with respect to its arguments \ $h_{\mu\nu}$ \ and \
$h^{\prime}_{\lambda\rho}$, which in what follows are operator
fields (40). Thus, here the problem of ordering the operator fields
to lowest order has been solved.

Now we can write out the following relations, which follow from the
exact quantum equations with the expansion indicated above. To
first order in \ $l_P$ \ we have
$$
\frac{1}{2}\,K^{(0)\,\lambda \rho}_{\mu \nu}\,h_{N\,\lambda \rho}=0
                                                     \eqno(46)
$$

We note that, using Eqs. (42), the operator (44) vanishes on the
quantity \ $(\xi_{\mu\,;\nu}+\xi_{\nu\,;\mu})$. Consequently, the
value of the operator (44) on the fields \ $h_{\mu\nu}$ \ and
$$
h^{\prime}_{\mu\nu}=h_{\mu\nu}+\xi_{\mu\,;\nu}+\xi_{\nu\,;\mu}   \eqno(47)
$$
coincide for any vector field \ $\xi_{\mu}$. This fact is a
consequence of the gauge invariance of the theory. Using the
indicated gauge invariance, any solution of Eq. (46) can be put
into the form
$$
\nabla^{(0)}_{\nu}\,h^{\nu}_{\mu}-\frac{1}{2}\,
\nabla^{(0)}_{\mu}\,h^{\nu}_{\nu}=0   \eqno(48)
$$
In what follows, we shall assume that the field satisfies the gauge
condition (48), which is convenient in a number of problems. It is
obvious that taking account of the gauge condition (48) the terms
in round brackets in operator (44) vanishes.

To clarify the question of the normalization of the gravitational
modes, we shall employ the following technique. The equation of
motion (46) can be obtained with the help of the action
$$
S^{(2)}=\int\,d^4x\,\sqrt{-g^{(0)}}\,
h^{\mu\nu}\,K^{(0)\,\lambda\rho}_{\mu\nu}\,h_{\lambda\rho}
  \eqno(49)
$$
Hence follows the canonically-conjugate momentum for the field
\ $h_{\mu\nu}$ \ and the simultaneous commutation  relations:
$$
\pi^{\mu\nu}=\sqrt{-g^{(0)}}\,\nabla^{(0) 0}\,h^{\mu\nu}\,,
$$
$$
[h_{\mu\nu}(x),\,\pi^{\lambda\rho}(y)\,]=
i\,\delta^{\lambda}_{(\mu}\,\delta^{\rho}_{\nu )}\,\delta^{(3)}(x-y)
    \eqno(50)
$$
Evidently, in Eq. (50) the fields are free of constraints (48).
Let's represent the field \ $h_{\mu\nu}$ \ in the form (compare
with the first term in Eq. (40))
$$
h_{\mu\nu}(x)=\sum_N\,\bigl(h_{N\,\mu\nu}(x)\,c_N+
h^*_{N\,\mu\nu}(x)\,c^{\dag}_N\bigr)   \eqno(51)
$$
The set of operators \ $\{c_N,\,c^{\dag}_N \}$ \ form a Heisenberg
algebra (1), and the functions \ $\{h_{N\,\mu\nu}\}$ \ satisfy Eq.
(46). Equations (50) and (51) lead to the following relations
reflecting the orthonormal nature of the set of the modes:
$$
i\,\int_{\Sigma^{(3)}}\,d^3x\,\sqrt{-g^{(0)}}\,
\bigl[h^{\mu\nu *}_M\,\nabla^{(0)\,0}\,h_{N\,\mu\nu}-
(\nabla^{(0)\,0}\,h_M^{\mu\nu\,*})\,h_{N\,\mu\nu}\,\bigr]=
\delta_{M,\,N}  \eqno(52)
$$
In the latter equations the integration extends over any spacelike
hypersurface \ $\Sigma^{(3)}$. As a result of Eqs. (46), integrals
(52) indeed do not depend on the hypersurface. It is natural to
assume that the gravitational modes satisfy conditions (52). The
significance of Eqs. (52) is that the normalization of the
coefficient functions in expansion (40) is given with its help.

In the second order in $l_P$, we obtain the following equations:
$$
\frac{1}{2}\,K^{(0)\,\lambda \rho}_{\mu \nu}\,h_{N_1 N_2\,\lambda \rho}
=-R^{(0)(2)}_{\mu \nu}\,(h_{N_1},\,h_{N_2})\,\,,        \eqno(53)
$$
$$
\frac{1}{2}\,K^{(0)\,\lambda \rho}_{\mu \nu}\,
h_{N_1 \mid N_2\,\lambda \rho}=
-2\,R^{(0)(2)}_{\mu \nu}\,(h_{N_1}^*,\,h_{N_2})\,\,,      \eqno(54)
$$
$$
\frac{1}{2}\,K^{(0)\,\lambda \rho}_{\mu \nu}\,
h^{F(\pm\pm)}_{N_1\,N_2\,\lambda \rho}=\pm\frac{i}{4}\,
\Bigl(\overline{\psi}^{(0)(\pm)}_{N_1}\,\gamma^c\,e^{(0)}_{c(\mu}\,
D^{(0)}_{\nu)}\,\psi^{(0)(\pm)}_{N_2}-
$$
$$
-e^{(0)}_{c(\mu}\overline{D_{\nu )}^{(0)}\,
\psi^{(0)(\pm)}_{N_1}}\,\gamma^c\,\psi^{(0)(\pm)}_{N_2}\,\Bigr)\,,
\eqno(55)
$$
$$
\frac{1}{2}\,K^{(0)\,\lambda \rho}_{\mu \nu}\,
h^{F(+-)}_{N_1\,N_2\,\lambda \rho}=\frac{i}{4}\,
\Bigl(\overline{\psi}^{(0)(+)}_{N_1}\,\gamma^c\,e^{(0)}_{c(\mu}\,
D^{(0)}_{\nu)}\,\psi^{(0)(-)}_{N_2}-
$$
$$
-e^{(0)}_{c(\mu}\overline{D_{\nu )}^{(0)}\,
\psi^{(0)(+)}_{N_1}}\,\gamma^c\,\psi^{(0)(-)}_{N_2}\,\Bigr)\,,
\eqno(56)
$$
$$
\frac{1}{2}\,K^{(0)\,\lambda \rho}_{\mu \nu}\,g^{(2)}_{(cl)\,\lambda \rho}=
-\sum_{|N|<N_0}\,R^{(0)(2)}_{\mu \nu}\,(h_N^*,\,h_N)+
$$
$$
+\frac{i}{4}\,\sum_{|N|<N_F}\,\Bigl(\overline{\psi}^{(0)(-)}_{N}\,\gamma^c\,
 e^{(0)}_{c(\mu}\,
D^{(0)}_{\nu)}\,\psi^{(0)(-)}_{N}- e^{(0)}_{c(\mu}\overline{{\cal
D}_{\nu )}^{(0)}\,
\psi^{(0)(-)}_{N}}\,\gamma^c\,\psi^{(0)(-)}_{N}\,\Bigr)\,.
\eqno(57)
$$
It is evident from Eq. (29) that torsion appears in the same order
\ $\sim l^2_P$. Here, however, we do not write out the
corresponding corrections for the connection.

We shall now briefly summarize the results obtained.

According to the dynamic quantization method, the quantization of
gravity starts with finding a solution of the classical
microscopic field equations of motion (for example, the solution
of Eqs. (42) in the example considered above). The classical
solution is determined by (or determines) the topology of
space-time. Then, using the classical approach, Eqs. (43) and
(46), which determine the single-particle modes
$\{\,\psi^{(\pm)}_N\,,\,h_{N\,\mu \nu}\}$, are solved. To solve
Eq. (46) the gauge must be fixed, since the operator (44) is
degenerate because of the gauge invariance of the theory. At the
first step these modes are determined in the zeroth approximation
according to the Planck scale, and their normalization is fixed
using Eqs. (39) and (52). Given the set of modes
$\{\,\psi^{(0)(\pm)}_N\,,\,h_{N\,\mu \nu}\}$, we can explicitly
write out the right-hand sides of Eqs. (53)-(57) and then solve
them for the two-particle modes $h_{N_1\,N_2\,\,\mu \nu}$,
$h_{N_1\,\mid N_2\,\,\mu \nu}$, and so on, and find the correction
$g^{(2)}_{(cl)\,\mu \nu}$ which is of second order in $l_P$ to the
classical component of the metric tensor. We call attention to the
fact that the right-hand side of Eq. (57) arises because the
operators must be normal-ordered. The solution of Eq. (57) can be
interpreted as a single-loop contribution to the average of the
metric tensor with respect to the ground state.

We note that if a nonsymmetric bilinear form were used on the
right-hand sides of Eqs. (53)-(57), then the condition that the
metric tensor be real would be violated. Consequently, the
condition that the metric tensor is real determines the ordering of
the operator fields in the equations of motion at least in second
order with respect to the operator fields.

It is important that all Eqs. (42), (46), and so on which arise are
generally covariant, since they are expansions of generally
covariant equations. Thus, the dynamic quantization method leads to
a regularized gauge-invariant theory of gravity, which contains an
arbitrary number of physical degrees of freedom.

We shall now make a remark about the compatibility of Eqs.
(53)-(57) and the analogous equations arising in higher orders. Let
$h_{\mu\nu}$ be an arbitrary symmetric tensor field and $K^{(0)}$
the operator (44), acting on this tensor field. It is easily
verified that, using Eqs. (42), we obtain the identity (compare
with Eq. (48))
$$
\nabla^{(0)}_{\nu}\,(K^{(0)}h)^{\nu}_{\mu}-
\frac{1}{2}\nabla^{(0)}_{\mu}\,(K^{(0)}h)^{\nu}_{\nu}=0\,.
\eqno(58)
$$
Consequently, in order for Eqs. (53)-(57) to be compatible the
right-hand sides of these equations must satisfy the same
identity. It is easy to see that this is indeed the case in lowest
order. Indeed, Eqs. (53)-(56) are identical to the analogous
classical equations arising when nonuniform modes (higher order
harmonics) and the subsequent expansion of the classical Einstein
equation in powers of the nonlinearity or the Planck lenght are
added to the uniform fields. Hence it follows that each term on
the right-hand sides of the "loop" equations of the type (57)
likewise satisfy the necessary identity, since these terms have
the same form as the right-hand sides of the "nonloop" Eqs.
(53)-(56).

In highest orders in creation and annihilation operators the
compatibility of arising equations follows from the gauge
invariance of the regularized Einstein equation. Indeed, the
identity (58) is the consequence of gauge invariance (invariance
relative to the general coordinate transformations) of the
equation. To clarify the quation let's rewrite the action (28)
(for simplicity with $m=0, \, \Lambda=0$) in the following form:
$$
S=S_g+S_{\psi}\,,   \eqno(59)
$$
$$
S_g=-\frac{1}{4l^2_P}\int\d^4
x\varepsilon_{abcd}\varepsilon^{\mu\nu\lambda\rho}e^a_{\mu}R^{bc}_{\nu\lambda}e^d_{\rho}\,,
\eqno(59 a)
$$
$$
S_{\psi}=\frac16\int\d^4
x\varepsilon_{abcd}\varepsilon^{\mu\nu\lambda\rho}\left[\frac{i}{2}\opsi
e^b_{\nu}e^c_{\lambda}e^d_{\rho}\gamma^aD_{\mu}\psi+h.c.\right]\equiv
\int\d^4 x\opsi\stackrel{\leftrightarrow}{\cal D}\psi\,. \eqno(59
b)
$$
Here $\stackrel{\leftrightarrow}{\cal D}$ is Dirac hermithian
operator, depending on other operator fields. The
Heisenberg--Dirac equations are written in the form
$$
\stackrel{\rightarrow}{\cal D}\psi=0\,, \qquad
\opsi\stackrel{\leftarrow}{\cal D}=0\,.   \eqno(60)
$$
In Eqs. (60) the disposition of creation and annihilation
operators is the same as in the action (59). Einstein equation is
the condition of stationarity of the action (59) relative to
variations of metric or tetrad. Evidently, the action (59) is
invariant under the general coordinate transformation even if the
fields are quantized. This follows from the facts that under the
coordinate transformations all fundamental fields transform
linearly and that the action (59) is a polynomial relative to the
fundamental fields. Therefore, if the material fields are on mass
shell (in our case this means that Eqs. (60) hold), the action
(59) is stationary under infinitesimal gauge transformation of
tetrad field only. This means that the quantum energy-momentum
tensor on mass shell  satisfies to some identity which in
classical limit transforms to the well known identity
$T^{\mu}_{\nu;\,\mu}=0$. From this quantum identity it follows
that if some quantum tetrad field satisfies Einstein equation,
then the field transformed by infinitesimal gauge transformation
also satisfies Einstein equation. From here the compatibility of
quantum Einstein equation follows, as well as the compatibility of
the chain of equations described above. However, this conclusion
is true only if quantum Dirac equations (60) hold, and the
operators in the action and energy-momentum tensor are placed so
as in Eq. (59). In other words, the creation and annihilation
operators in Eqs. (59) and (60) must be placed identically. This
is the guarantee of self-consistency of the chain of equations
arising from exact quantum Einstein and motion equations.

We also call attention to the fact that in the dynamic
quantization method it is implicitly assumed that the quantum
anomaly is absent in the algebra of the first class constraints
operators. Consequently, the dynamic quantization method must be
justified in each specific case by concrete calculations, which
must be not only mathematically correct but also physically
meaningful.

\centerline{ }
\centerline{ }
{\large\sl {4. The possible solution
of cosmological constant problem}}
\centerline{ }
\centerline{ }

It follows from Eqs. (43) that in lowest order the Dirac field
$$
 \psi^{(1)}(x)=\sum_{|N|<N_0}\left(a_N\psi^{(0)(+)}_N(x)+b^{\dag}_N\psi^{(0)(-)}_N(x)
\right)    \eqno(61)
$$
satisfies the Dirac equation
$$
(ie_a^{(0)\,\mu}\gamma^aD_{\mu}^{(0)}-m)\psi^{(1)}(x)=0\,.
\eqno(62)
$$
Here $D_{\mu}^{(0)}$ is the covariant derivative operator in
zeroth order:
$$
D_{\mu}^{(0)}=\partial/\partial
x^{\mu}+(1/2)\omega^{(0)}_{ab\mu}\sigma^{ab}+ieA^{(0)}_{\mu}\,,
   \eqno(63)
$$
and $A_{\mu}$ is the gauge field.

It follows from Eq. (62) that the charge
$$
Q=\int\d^3x\sqrt{-g^{(0)}}e^{(0)0}_a(\opsi^{(1)}\gamma^a\psi^{(1)})\,.
\eqno(64)
$$
conserves (compare with (39)).

According to (33) the contribution of the Dirac field to the
energy-momentum tensor in lowest order is equal to
$$
T^{(2)}_{\psi\,\mu\nu}=\Re[i\opsi^{(1)}\gamma^ae^{(0)}_{a(\mu}D_{\nu)}^{(0)}\psi^{(1)}]\,.
\eqno(65)
$$
Using Eqs. (36) it is easy to find vacuum expectation value of the
quantity (65):
$$
\langle
T^{(2)}_{\psi\,\mu\nu}\rangle_0=\Re\left[i\sum_{|N|<N_0}\opsi_N^{(0)(-)}\gamma^ae^{(0)}_{a(\mu}
D^{(0)}_{\nu)}\psi^{(0)(-)}_N\right]\,. \eqno(66)
$$

Now let us take into account that the scenario described by the
inflation theory is realized in Universe. It follows from here in
conjunction with the used quantization method that in zeroth
approximation the metric is expressed as
$$
\d s^{(0)\,2}=\d t^2-a^2(t)\,\d\Omega^2\,,  \eqno(67)
$$
where $\d\Omega^2$ is the metric on unite sphere $S^3$, and $a(t)$
is the scale factor of Universe at the running moment of time $t$.
It follows from (67) that $e_a^{(0)0}=\delta^0_a$ and
$\sqrt{-g^{(0)}}\d^3x=\d V^{(0)}(t)$, where $\d V^{(0)}(t)$ is the
volume element of 3-space in the running moment of time. From
conservation of operator (54) the conservation of the set of
integrals
$$
\int\d
V^{(0)}(t)\,\psi^{(0)(\pm)\dag}_N\psi^{(0)(\pm)}_M=\delta_{NM}\,.
\eqno(68)
$$
follows. The equality to unity of integrals (68) means that the
wave functions $\psi^{(0)(\pm)}_M$ are normalized relative to the
volume of all Universe, so that the charge operator has the form
$$
Q=\sum_{|N|<N_0}(a^{\dag}a_N+b_Nb_N^{\dag})\,.  \eqno(69)
$$

The idea how the vacuum expectation value of the matter
energy-momentum tensor becomes enough small at present is
demonstrated by the following estimation.

According to (68) we have:
$$
\left|\opsi_N^{(0)(\pm)}\psi_N^{(0)(\pm)}\right|\sim\frac{1}{a^3(t)}\,.
\eqno(70)
$$
Therefore the estimation for the value (66) is the following:
$$
\langle
T^{(2)}_{\psi\,\mu\nu}\rangle_0\sim\frac{N_0k_{max}}{a^3(t)}\,,
\eqno(71)
$$
where $k_{max}$ is the value of the order of maximal momentum of
the modes $\{\psi_N^{(0)(\pm)}\}$. It is naturally to suppose that
$$
k_{max}\sim l^{-1}_P\sim G^{-1/2}\sim 10^{33}{cm}^{-1}\,.
\eqno(72)
$$
Since the numerator in the right hand side of relation (71) is
finite and the denominator is proportional to the volume of
Universe which swells up approximately $10^{100}$ times more
according to inflation scenario, the quantity (71) can be found
enough small at present.

On the other hand, it is seen from the estimation (71) that at
early stages of Universe evolution the quantum fluctuations played
decisive role because the scale of the Universe were small.

One should pay the attention to the fact that the dynamics of the
system creates two opposite tendencies for mode frequencies
changing.

According to the first tendency the frequencies $\omega$ of all
one-particles modes change in time according to the low
$$
\omega\sim\frac{1}{a(t)}\,.   \eqno(73)
$$
The low (73) is valid in relativistic case. So, all frequencies
decrees with expansion of Universe.

Now let us write out the first items of formal solution of Dirac
equation (32) or (60) neglecting gravity degrees of freedom (i.e.
in the case of flat space-time) but in the presence of gauge
field:
$$
\psi(x)=\psi^{(1)}(x)+e\int\d^4y\,S_{ret}(x-y)\,A^{(1)}_{\mu}(y)
\gamma^{\mu}\psi^{(1)}(y)+
$$
$$
+4\pi\,e^2\int\!\!\int\d^4y\d^4z\,S_{ret}(x-y)\,D_{ret}(y-z)\left(\opsi^{(1)}(z)
\gamma_{\mu}\psi^{(1)}(z)\right)\gamma^{\mu}\psi^{(1)}(y)+\ldots\,,
\eqno(74)
$$
$$
\left(i\gamma^{\mu}\partial_{\mu}-m\right)S_{ret}(x)=\delta^{(4)}(x)\,,
\eqno(75)
$$
$$
\partial_{\mu}\partial^{\mu}D_{ret}(x)=\delta^{(4)}(x)\,.
\eqno(76)
$$
Here $S_{ret}(x)$ and $D_{ret}(x)$ are the retarded Green
functions satisfying Eqs. (75) and (76). It is seen from the
solution (74) that the exact field $\psi(x)$ have much more
nonzero Fourier components than the field $\psi^{(1)}$. Hence, the
exact solution of quantum Dirac equation has all Fourier
components despite the field of first approximation $\psi^{(1)}$
has Fourier components only with finite momenta. From here we see
the opposite dynamic tendency: the frequencies of modes
effectively increases as a consequence of interaction. If the fact
of strict conservation of the charge is taken into account
\footnote{It means strict conservation of quantum charge operator
which in lowest order transforms into expression (64)}, the
conclusion about noncompact "packing" of modes in momentum space
should be made. Indeed, let's calculate the mean value of charge
operator relative to a state $|\,\rangle$. We have:
$$
\int\d^3x\langle
\,|\,\psi^{\dag}(x)\psi(x)\,|\;\rangle=\int\frac{\d^3k}{(2\,\pi)^3}
\langle \,|\,\psi^{\dag}_{|\bk|}\psi_{|\bk|}\,|\;\rangle=\const\,,
$$
$$
\psi_{|\bk|}=\int\d^3x\,e^{-i\bk\bx}\psi(x)\,.
$$
The last relation means also that the integral
$$
\int\d^3x\,\tr\langle
\,|\,T\,\psi(x)\opsi(y)\,|\;\rangle\,\gamma^0\,,
$$
$$
y\longrightarrow x\,, \qquad y^0>x^0
$$
constructed with the help of correlator $\langle
\,|\,T\,\psi(x)\opsi(y)\,|\;\rangle$ is conserved in time. But the
mean value of energy-momentum tensor is constructed with the help
of the same correlator. From here it is seen the effect of
"loosening of mode packing" in momentum space. This effect is
absent in the theory with dense packing of modes in all diapason
of momenta since all states in momentum space are filled by the
corresponding modes.

To solve the problem of "loosening of mode packing" in momentum
space one must solve quantum kinetic equation for state density in
momentum space. This problem is not solved in this work. But the
fact of noncompact "packing" of modes in momentum space plays an
important role in our consideration. Thus, the noncompact
"packing" of modes in momentum space is taken here as an
assumption.

At a dense "packing" of modes in momentum space the neighbouring
momenta differ by the quantity of the order of $\Delta k_{min}\sim
1/a(t)$. Therefore
$$
\d N\sim\frac{a^3(t)\d^3k}{(2\,\pi)^3}\,.   \eqno(77)
$$
At noncompact "packing" of modes in momentum space the
neighbouring momenta differ by the greater quantity. Assume that
at small momenta this difference at present is of the order of
$\Delta k_{min}\sim 1/\lambda_{\max}$. Furthermore, we shall use
Lorentz-invariant measure in momentum space $[\d^3k/|\bk|]$. Thus
we obtain instead of (77) the following estimation for the total
number of physical degrees of freedom:
$$
N_0\sim\lambda^3_{max}\int^{k_{max}}\frac{\d^3k}{(2\pi)^3(\lambda_{max}|\bk|)}\sim
(\lambda_{max}k_{max})^2\,.   \eqno(78)
$$
Now using (71) and (78) we find:
$$
16\pi G\langle
T_{\mu\nu}\rangle_0\sim\frac{l_P^2\,\lambda^2_{max}\,k^3_{max}}{a^3(t)}\leq\Lambda\,,
\eqno(79)
$$
and from here
$$
a(t_0)\geq\frac{(l_P\,\lambda_{max})^{2/3}\,k_{max}}{\Lambda^{1/3}}\,.
\eqno(80)
$$
If one assume that
$$
\lambda_{max}\sim 10^{24}cm\sim 10^{-4}L\,,  \eqno(81)
$$
where $L=10^{28}cm$ (the dimension of observed part of Universe),
then with the help of relations (2), (4), (81) and (80) we find
the following estimation for the present dimension of Universe:
$$
a(t_0)\geq 10^{17}L\,.   \eqno(82)
$$

At obtaining the estimation (82) it was assumed that the
fundamental field theory is not supersymmetric. If one assume that
the fundamental theory is supersymmetric, but the spontaneous
breaking of supersymmetry occurs on the momentum $\sim k_{SS}$,
then the estimation of the dimension of Universe is changed.
Indeed, in this case instead of (72) we have
$$
k_{max}\sim k_{SS}\,,  \eqno(83)
$$
since according to (9) and (10) the boson and fermion
contributions to the vacuum expectation value of energy-momentum
tensor with momenta greater than $k_{SS}$ are mutually cancelled.
Therefore instead of (80) we obtain:
$$
a(t_0)\geq\frac{(l_P\,\lambda_{max})^{2/3}\,k_{SS}}{\Lambda^{1/3}}
\sim10^{29}cm\sim10L\,. \eqno(84)
$$
At obtaining the numerical estimation of right hand side Eq. (84)
we used assumptions (81) and the popular assumption in particle
physics that $k_{SS}\sim 10^3 GeV\sim 10^{17}cm^{-1}$.

The inclusion of quantum fluctuations of others fields into our
estimations does not changes the result. This is clear already
from dimensional considerations.

The inclusion of higher order corrections by perturbation theory
also does not changes the obtained estimations. Indeed, all known
fundamental interactions except for gravitational are
renormalizable and thus can be considered by perturbation theory
without changing fundamental properties of the vacuum. But the
gravitational quantum corrections are obtained by expanding in
Planck scale $l_P$. Again from dimensional considerations it is
clear that such corrections at passing to the following order in
our theory have the comparative value
$$
\sim\left(\frac{l_P}{a(t_0)}\right)^2N_0\sim\left(\frac{l_P}{a(t_0)}\right)^2(\lambda_{max}
k_{max})^2\leq\left(\frac{\lambda_{max}}{a(t_0)}\right)^2\ll 1\,.
\eqno(85)
$$

\centerline{ }
\centerline{ }
{\large\sl {5. Discussion}}
\centerline{ }
\centerline{ }

Does the Casimir effect survives in proposed theory? The answer to
this question is positive. Indeed, the attraction force between
plates of condenser which is caused by Casimir effect is the
derivative of sum of photon zero-point energies with respect to
distance between plates. But only modes with wavelength
commensurable with the distance between plates $d$ really give the
contribution in this derivative. And since $d\ll\lambda_{max}$ the
distortion of Casimir effect does not occurs.

One should pay attention to the fact that in presented theory with
noncompact packing of modes the gravitational and gauge
interaction forces does not become weaker. It is seen from quantum
equations of motion which have the canonical form with usual
interaction constants. Thus the interaction between any modes has
the usual strength.

Further, one can assume that in considered theory the
stochastization of phases of modes takes place on distances less
than $\lambda_{max}$. Under the phase stochastization we mean that
any correlation between phases of wave packets spased by an enough
distance can not take place. Such stochastization must occur if
considered theory is the long-wavelength limit of discrete quantum
theory of gravity discussed in [12]. The point is that in
long-wavelength limit the lattice action $S$ transforms to the
action which is expressed as follows:
$$
S=S_{Einstein}+\Delta S\,.
$$
Here $S_{Einstein}$ is standard Einstein action which does not
retains any information about the structure of lattice, and
$\Delta S$ depends only on higher derivatives of the fields and
also it essentially depends on the structure of the lattice.
Therefore equations of motion contain the items with higher
derivatives of fields and casual coefficients depending on
structure of irregular lattice (simplicial complex). These items
play negligible part for low frequencies modes but their part
increase with increasing of mode frequency. The items with higher
derivatives of fields and casual coefficients lead to diffusional
propagation of modes and so to stochastization of phase on large
distances. But just due to this circumstance the high frequency
wave packets can be localized in relatively small regions of
space. This means that noncompact "packing" of modes in momentum
space does not affects to the possibility of localization of high
frequency wave packets.

Let's consider, for example, the system of finite number of
electrons and positrons with wavelengths much less than
$\lambda_{max}$. Assume that we are interested in the usual
problem of particle physics: the scattering matrix problem. The
dynamics of real relativistic particles is described by the usual
Dirac and Maxwell equations. The dynamic process of particles
localized in finite space volume $v\ll\lambda^3_{max}$ is studied.
Since the matrix elements between localized states and
nonlocalized states tends to zero as $a^{-3/2}(t_0)$, so only
matrix elements between localized states are significant in the
studied problem. This conclusion is true also with respect to
virtual modes. From here it follows that for description of
processes proceeding in finite volume of space $v$, one must use
the renormalized quantum fields $(\psi_r,\,\ldots)$ which are
normalized to the volume $v$. This means that the wave functions
of the states $\{\psi_{r\,N}(x),\ldots\}$ which create and
annihilate the renormalized fields
 are normalized to
the volume $v$, these wave functions form the complete set of one
particle wave functions with confined energies, and the
corresponding creation and annihilation operators satisfy to
standard relations (36). Note that the quantization conditions
(36), i.e. nullification of ground state by annihilation
operators, follow from the fact that the causal correlators
$\langle 0\,|\,T\,\psi(x)\,\opsi(y)\,|0\,\rangle$ describe
propagation omly positive-frequency waves. It seems that more
general and correct definition of ground state $|0\rangle$ instead
of definition (26) or (36) is that the amplitudes
$$
\langle 0\,|\,T\,\psi(x)\,\opsi(y)\,|\,0\,\rangle\,, \qquad
\langle 0\,|\,T\,A_i(x)\,A_j(y)\,|0\,\rangle\,,\,\ldots  \eqno(86)
$$
describe the propagation of only positive-frequency waves if the
times $x^0$ and $y^0$ are close to some time moment $t_0$. Again
the definition of vacuum depends on the moment of time $t_0$. At
present the state of Universe is close to the ground state. One
can say that the renormalized fields $(\psi_r,\,\ldots)$ are the
secondary quantized fields with the complete (at confined
energies) and normalized on volume $v$ set of one-particle states
$\{\psi_{r\,N}(x),\ldots\}$. Thus the cosmological fields
$(\psi,\,\ldots)$ from which quantum global Einstein equation is
composed and the secondary quantized fields $(\psi_r,\,\ldots)$
are different though they describe the particles with the same
quantum numbers. The causal correlators constructed from
renormalized fields $\psi_r$ (renormalized correlators) and thus
describing local interactions also satisfy the conditions (86).
Since local states normalized to volume $v$ have compact "packing"
in momentum space (at least for experimentally tested momenta),
the renormalized correlators satisfy the standard equations:
$$
(i\gamma^{\mu}\partial_{\mu}-m)\,\langle
0\,|\,T\,\psi(x)\,\opsi(y)\,|\,0\,\rangle=i\,\delta^4(x-y)\,,
\ldots\,,
$$
which are true at $|x^0-y^0|\gg l_P, \ |\bx-\by|\gg l_P$. And
since the calculations of $S$-matrix elements are performed by
using the standard Dirac and Maxwell equations with usual value of
charge and others parameters, as a result the usual expressions
for $S$-matrix elements are obtained.

Let's emphasize that at solving cosmological problems the retarded
Green functions are used but at calculating $S$-matrix elements
the causal or Feynman one are used.

We make the last remark about violation of Lorentz invariance in
the theory. Since as a matter of fact the regularization is
performed here by energy but not Lorentz invariant square of
4-momentum, so Lorentz invariance can be violated. However, until
the processes with low energies (in comparison with the cutoff
energy) are studied the violation of Lorentz invariance is
negligible. The regularization by energy of calculations in QED is
used, for example, in [13], and at the same time Lorentz
invariance is not violated at low energies. Therefore the fact
that all observed phenomena in nature are Lorentz covariant does
not contradicts to the proposed theory since these phenomena has
been observed at confined energies.

Despite many remaining unclear questions, it seems to us that
presented here ideas worth to be discussed.  \\

\centerline{ }

ACKNOWLEDGMENT

I thank participants of seminar of prof. A. A. Belavin for useful
discussion, and especially prof. B. G. Zakharov.  This work was
supported by the Program for Support of Leading Scientific Schools
$\sharp$ 2044.2003.2. and RFBR $\sharp$ 04-02-16970-a.

\centerline{ }
\centerline{ }
{\large\sl {Appendix}}
\centerline{}
\centerline{ }

For clearness it is useful to see the phenomenon of imposing the
second class constraints without changing of quantum field
equations on an example of free Klein--Gordon theory. The
Klein--Gordon fields are expanded as follows:
$$
\phi(x)=\sum_{\bk}\frac{1}{\sqrt{2\omega_{\bk}}}\left(
a_{\bk}\phi_{\bk}(x)+a_{\bk}^{\dag}\phi_{\bk}^*(x)\right)\,,
$$
$$
\pi(x)=-i\sum_{\bk}\sqrt{\frac{\omega_{\bk}}{2}}\left(
a_{\bk}\phi_{\bk}(x)-a_{\bk}^{\dag}\phi_{\bk}^*(x)\right)\,,
$$
$$
\omega_{\bk}=\sqrt{\bk^2+m^2}\,, \qquad
[a_{\bk},\,a_{\bp}^{\dag}]=\delta_{\bk\bp}\,.   \eqno(A\,1)
$$
Here $\{\phi_{\bk}(x)\}$ is the complete set of orthonormal
funcrions, so that
$$
\sum_{\bk}\phi_{\bk}(x)\,\phi_{\bk}^*(y)\bigg|_{x^0=y^0}=\delta^{(3)}(\bx-\by)\,,
$$
$$
\Delta\,\phi_{\bk}(x)=-\bk^2\,\phi_{\bk}(x)\,.   \eqno(A\,2)
$$
The Hamiltonian
$$
{\cal
H}=\int\d^3x\left(\frac12\pi^2+\frac12\nabla\phi\nabla\phi+\frac{m^2}{2}\phi^2\right)=
\frac12\sum_{\bk}\omega_{\bk}\left(a_{\bk}a_{\bk}^{\dag}+a_{\bk}^{\dag}a_{\bk}\right)\,.
\eqno(A\,3)
$$
Equations of motions are obtained with the help of Eqs.
(A\,1)-(A\,3):
$$
\dot{\phi}(x)=-i[\phi(x),\,{\cal
H}]=-i\sum_{\bk}\sqrt{\frac{\omega_{\bk}}{2}}\left(
a_{\bk}\phi_{\bk}(x)-a_{\bk}^{\dag}\phi_{\bk}^*(x)\right)=\pi(x)\,,
$$
$$
\ddot{\phi}(x)=\dot{\pi}(x)=-i[\pi(x),\,{\cal H}]=
$$
$$
=-\sum_{\bk}\frac{\omega_{\bk}^2}{\sqrt{2\omega_{\bk}}}\left(
a_{\bk}\phi_{\bk}(x)+a_{\bk}^{\dag}\phi_{\bk}^*(x)\right)=(\Delta-m^2)\,\phi(x)\,.
\eqno(A\,4)
$$

Now let us impose any number of pairs of second class constraints
$$
a_{\bk_i}=0\,, \qquad a_{\bk_i}^{\dag}=0\,, \qquad
i=1,\,2\,\ldots\,.  \eqno(A\,5)
$$
Then the sums $\sum_{\bk}$ in  $(A\,1)$, $(A\,3)$ and $(A\,4)$
transform to the reduced sums $\sum_{\bk\neq\bk_i}$. Nevertheless
equation of motion $(A\,4)$ retains its canonical form
$$
\left(\partial^2/(\partial x^0)^2-\Delta+m^2\right)\,\phi(x)=0\,.
\eqno(A\,5)
$$
The dynamical reason for this conclusion in considered example is
that the commutators of the constraints $(A\,5)$ with Hamiltonian
are proportional to the constraints, i.e. they are equal to zero
in a weak sense:
$$
[a_{\bk_i},\,{\cal H}]=\omega_{\bk_i}a_{\bk_i}\,, \qquad
[a_{\bk_i}^{\dag},\,{\cal H}]=-\omega_{\bk_i}a_{\bk_i}^{\dag}\,.
\eqno(A\,7)
$$
Therefore, as it was shown in Section 2, equations of motion in
reduced theory must retain their canonical form.

\centerline{ }
\centerline{REFERENCES}
\centerline{ }
\centerline{ }

\begin{itemize}
\item[1. ] S.N. Vergeles, Theoretical and Mathematical Physics,
{\bf {112}}, 899 (1997).
\end{itemize}
\begin{itemize}
\item[2. ] S.N. Vergeles, Zh. Eksp. Teor. Fiz. {\bf {118}}, 996
(2000) [JETP. {\bf {91}}, 859 (2000)].
\end{itemize}
\begin{itemize}
\item[3. ] S.N. Vergeles, Zh. Eksp. Teor. Fiz. {\bf {113}}, 1566
(1998) [JETP {\bf {86}}, 854 (1998)].
\end{itemize}
\begin{itemize}
\item[4. ] S.N. Vergeles, Zh. Eksp. Teor. Fiz. {\bf {117}}, 5
(2000) [JETP {\bf {90}}, 1 (2000)].
\end{itemize}
\begin{itemize}
\item[5. ] E. Benedict, R. Jackiw, H.-J. Lee, Phys.Rev. {\bf
{D54}},6213(1996); D. Cangemi, R. Jackiw, B. Zwiebach, Ann. Phys.
(N.Y.) {\bf {245}},408(1996); D. Gangemi and R. Jackiw, Phys.Lett.
{\bf {B337}}, 271(1994); Phys.Rev. {\bf {D50}}, 3913(1994); D.
Amati, S. Elitzur and E. Rabinovici, Nucl.Phys. {\bf {B418}},
45(1994); D. Louis-Martinez, J Gegenberg and G. Kunstatter,
Phys.Lett. {\bf {B321}}, 193(1994); E. Benedict, Phys.Lett. {\bf
{B340}}, 43(1994); T. Strobl, Phys.Rev. {\bf {D50}}, 7346(1994).
\end{itemize}
\begin{itemize}
\item[6. ] S. Weinberg, Rev. Mod. Phys. {\bf{61}},1 (1989).
\end{itemize}
\begin{itemize}
\item[7. ] H. B. G. Casimir, Proc. K. Ned. Akad. Wet. {\bf{51}},
635 (1948).
\end{itemize}
\begin{itemize}
\item[8. ] M. J. Sparnaay, Nature (London), {\bf{180}},334(1957).
\end{itemize}
\begin{itemize}
\item[9. ] B. S. DeWitt, Phys. Reports, {\bf{19 C}}, 295 (1975).
\end{itemize}
\begin{itemize}
\item[10. ] B. Zumino, Nucl. Phys.{\bf{B 89}},535 (1975).
\end{itemize}
\begin{itemize}
\item[11. ] G. E. Volovik, E-print archives gr-qc/0101111; "The
Universe in a
 Helium Droplet", Clarendon Press. Oxford 2003.
\end{itemize}
\begin{itemize}
\item[12. ] S. N. Vergeles, Zh. Eksp. Teor. Fiz. {\bf {124}}, 1203
(2003).
\end{itemize}
\begin{itemize}
\item[13. ] P.A.M.Dirac, Lectures on quantum field theory, Yeshiva
University, New York. 1967.
\end{itemize}
\begin{itemize}
\item[14. ] S. Randjbar-Daemi, E. Sezgin, E-print archives
hep-th/0402217; Robert H. Brandenberger, S. Randjbar-Daemi,E-print
archives hep-th/0404228.
\end{itemize}
\begin{itemize}
\item[15. ] T. Padmanabhan, E-print archives gr-qc/0204020;
hep-th/0212290.
\end{itemize}

\end{document}